\documentclass{article}
\usepackage{spconf,amsmath,graphicx,amssymb}
\usepackage{caption,subcaption}
\usepackage{url}
\usepackage{setspace}
\usepackage[dvipsnames]{xcolor}

\title{A MULTI-VIEW APPROACH TO AUDIO-VISUAL SPEAKER VERIFICATION}
\name{Leda Sar{\i}\textsuperscript{1,2}\thanks{This work was done when L. Sar{\i} was an intern at Facebook.}, Kritika Singh\textsuperscript{1}, Jiatong Zhou\textsuperscript{1}, Lorenzo Torresani\textsuperscript{1}, Nayan Singhal\textsuperscript{1},  Yatharth Saraf\textsuperscript{1}}
\address{\textsuperscript{1}Facebook AI, USA \\ \textsuperscript{2}Department of Electrical and Computer Engineering, University of Illinois at Urbana-Champaign, USA}
\begin{document}
%
\maketitle
\begin{abstract}
Although speaker verification has conventionally been an audio-only task, some practical applications provide both audio and visual streams of input. In these cases, the visual stream provides complementary information and can often be leveraged in conjunction with the acoustics of speech to improve verification performance. In this study, we explore audio-visual approaches to speaker verification, starting with standard fusion techniques to learn joint audio-visual (AV) embeddings, and then propose a novel approach to handle cross-modal verification at test time. Specifically, we investigate unimodal and concatenation based AV fusion and report the lowest AV equal error rate (EER) of 0.7\% on the VoxCeleb1 dataset using our best system. As these methods lack the ability to do cross-modal verification, we introduce a multi-view model which uses a shared classifier to map audio and video into the same space. This new approach achieves 28\% EER on VoxCeleb1 in the challenging testing condition of cross-modal verification.
\end{abstract}
\section{Introduction}
\label{sec:intro}
Speaker recognition and verification systems are conventionally based on the speech component as speech is a medium that partially represents the identity of the speaker. However, in a noisy acoustic environment, it can become harder to distinguish different speakers based only on speech signals. In such cases, humans often rely on other signals for the identity which are not affected by acoustic noise, such as facial features. Because of the complementary nature of audio and video, several audio-visual (AV) systems have been proposed for speaker identification \cite{wu2006multi,senior1999use,shon2019noise}. 


Such AV identification systems vary depending on their fusion strategies and modeling approaches. As described in other multimodal studies, e.g. \cite{katsaggelos2015audiovisual}, fusion methods include early, mid-level and late fusion. Early fusion concatenates inputs and learns joint features of both modalities, mid-level fusion combines information after some independent processing of the two modalities and late fusion mainly consists of score fusion from unimodal systems. As for modeling approaches, earlier systems make use of probabilistic models such as dynamic Bayesian networks~\cite{wu2006multi} whereas recent studies focus on neural network based modeling \cite{shon2019noise}.  

A common approach to achieve speaker verification is to extract a speaker representative embedding from the given utterance and compare a pair of embeddings using a distance measure to determine if the given utterances belong to the same person. Earlier studies have used i-vectors \cite{dehak2010front}, as speaker representation and probabilistic linear discriminant analysis (PLDA) scoring for verification. In recent studies, neural network based speaker embeddings, such as x-vectors~\cite{snyder2017deep}, are used. These systems usually process the input speech by a network that generates a sequence of features for the utterance which are then aggregated into a single vector to represent the speaker embedding \cite{xie2019utterance,yadav2020frequency}. 
These aggregation or summarization methods range from temporal pooling to cluster based approaches which are also used in computer vision studies such as NetVLAD \cite{arandjelovic2016netvlad} and GhostVLAD \cite{zhong2018ghostvlad,xie2019utterance}. 

In this study, we first investigate AV speaker verification by learning AV speaker embeddings which assumes that audio and video data are simultaneously available at test time. To our knowledge, we achieve the best reported AV speaker verification performance on the VoxCeleb1 \cite{nagrani2017voxceleb} and VoxCeleb2 \cite{Nagrani18} datasets. These AV systems assume the availability of both modalities during test time. However, in many practical settings, one of the modalities is degraded or may be missing on one side of the verification pair. For instance, the speaker may be off-screen or have their camera disabled while they are actively speaking in which case only the audio stream is usable for verification. The audio may be missing or corrupted in some other scenarios. There may also be some verification pairs where audio is the usable modality on one side and video on the other. Here we need to do cross-modal matching, i.e. verifying if a video and an audio signal represent the same person. Late fusion or mid-level fusion cannot handle such cases. 
Therefore, we propose a multiview approach that allows us to perform verification in the case where a pair does not have matching modalities available, either audio or video. The proposed multiview approach achieves this by mapping audio and video into the same space by using a shared classifier on top of the unimodal encoders. 


\section{Related Work}
\label{sec:prev}
There have been several studies on VoxCeleb1 and VoxCeleb2 datasets and several benchmarks have been published. The typical setup is to train the system on VoxCeleb2-dev set and test on VoxCeleb1-test verification pairs.
These setups differ in network architecture and the embedding aggregation steps. For example, in \cite{Nagrani18}, ResNet-50 architecture and time average pooling is used which achieves 4.2\% EER. In \cite{xie2019utterance}, thin ResNet-34 architecture is used along with GhostVLAD pooling that results in 3.2\% EER. 
In \cite{yadav2020frequency}, a convolutional attention model is proposed for time and frequency dimensions and GhostVLAD based aggregation is applied and the model achieves 2.0\%. 
The lowest EER on the VoxCeleb1 test set is reported in \cite{zeinali2019but}, where their best system makes use of data augmentation and system combination. 

Although the VoxCeleb2 dataset comes with videos, there are only a few studies on audio-visual approaches for speaker verification. In \cite{shon2019noise}, pretrained face and voice embeddings are fused using a cross-modal attention mechanism on short speech segments (0.1s or 1s). In their tests on VoxCeleb2, they obtain an EER of 5.3\%. They also analyze the performance in the case of noisy or missing modality and their performance degrades to 7.7\% and 12.2\% when voice and face embeddings are omitted, respectively. In \cite{nagrani2020disentangled}, an audio-visual self-supervised approach is used to train a system that learns identity and context embeddings separately. As a comparison, they also report audio-only fully-supervised training results on the VoxCeleb1 test set which achieves 7.3\% EER. Since they use only 20\% of the VoxCeleb2 dataset for training and there is not a standard set of verification pairs for VoxCeleb2, their AV results are not directly comparable to the previous study.   

Cross-modal processing has been recently used in different combinations such as audio-video \cite{nagrani2018learnable,nagrani2020disentangled,tao2020audio, nawaz2019deep} and speech-text \cite{sari2020training}. The common approach in these studies is to map inputs from different modalities into a shared space to achieve cross-modal retrieval. 
For example, in \cite{nagrani2018learnable}, contrastive loss is used to learn to map matching face and voice embeddings to the same space. In \cite{tao2020audio}, same-different classification is performed on the cosine scores between face and voice embeddings to train the system. In \cite{nawaz2019deep}, a novel loss function is proposed to learn the embeddings in a shared space. Their loss function tries preserving neighborhood constraints within and across modalities. 

\begin{figure*}[t]
    \centering
    \begin{subfigure}[b]{0.2\textwidth}
         \centering
         \includegraphics[scale=0.5]{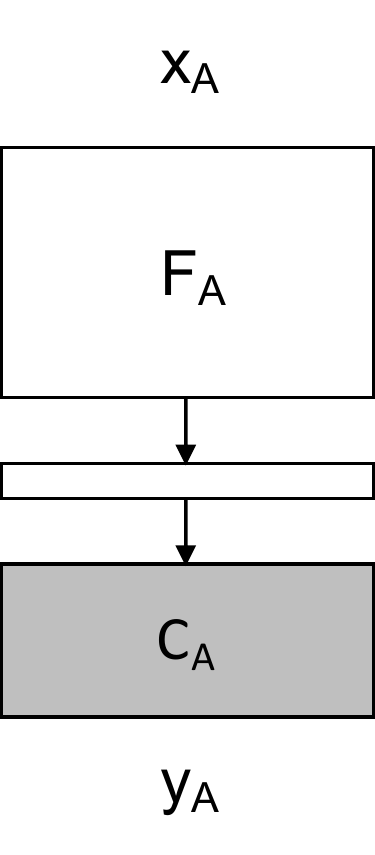}
         \caption{Unimodal: Audio-only}
         \label{fig:unimodala}
     \end{subfigure}
     \hspace{-5pt}
     \begin{subfigure}[b]{0.2\textwidth}
         \centering
         \includegraphics[scale=0.5]{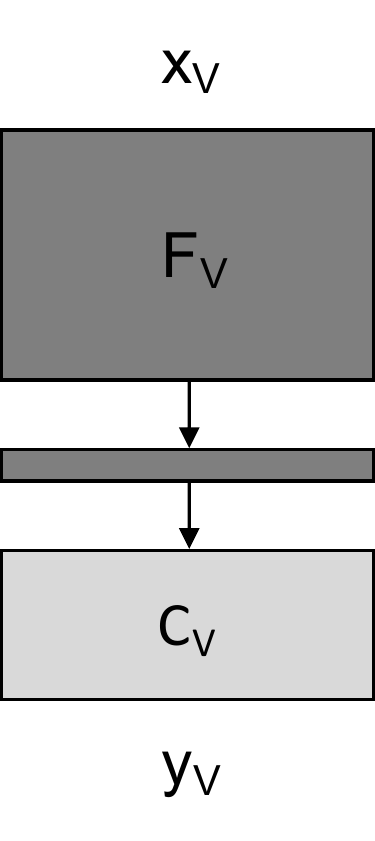}
         \caption{Unimodal: Video-only}
         \label{fig:unimodalv}
     \end{subfigure}
     \hfill
     \begin{subfigure}[b]{0.25\textwidth}
         \centering
         \includegraphics[scale=0.5]{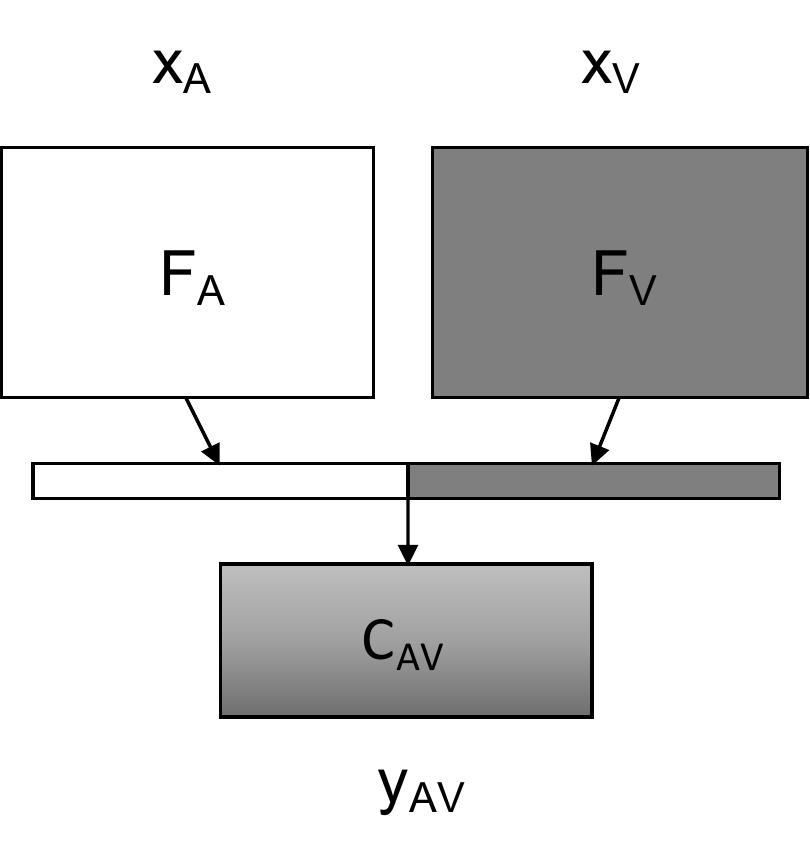}
         \caption{Mid-level AV fusion}
         \label{fig:concat}
     \end{subfigure}
     \hfill
     \begin{subfigure}[b]{0.25\textwidth}
         \centering
         \includegraphics[scale=0.5]{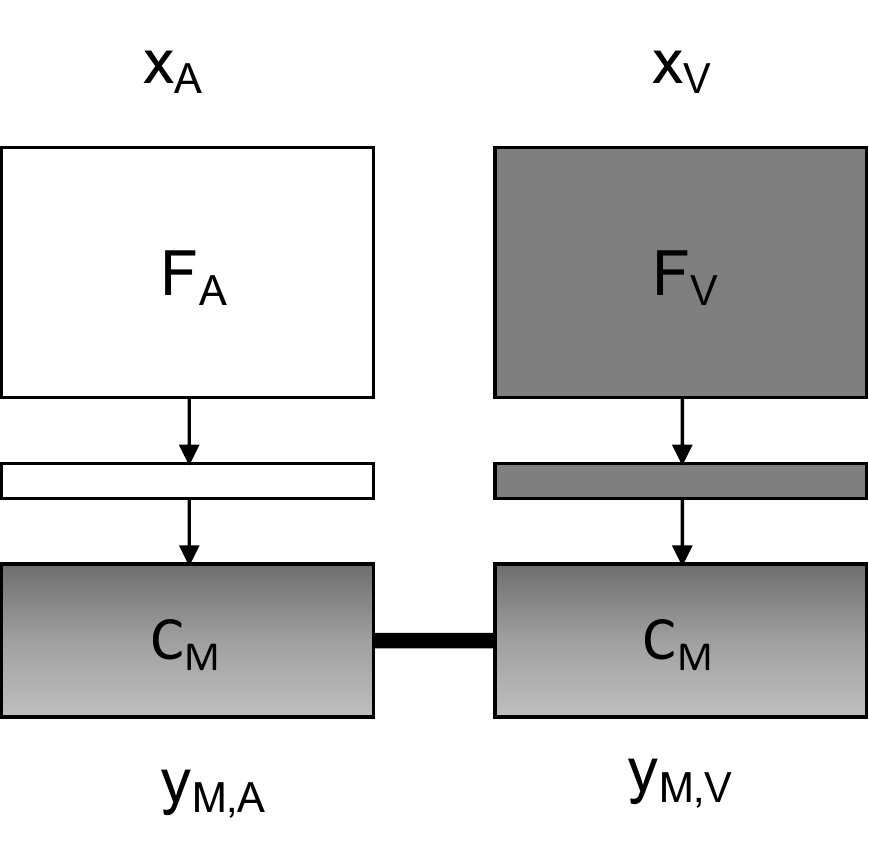}
         \caption{Multi-view}
         \label{fig:multiview}
     \end{subfigure}
    \caption{Encoder and classifier structures of various models: (a) A-only unimodal system; (b) V-only unimodal system; (c) Mid-level AV fusion model; (d) The multi-view model.}
    \label{fig:flowcharts}
    \vspace{-10pt}
\end{figure*}

\section{Unimodal and Multimodal Models}
\label{sec:uni-multi}
In the training stage, our verification models are optimized to learn speaker discriminative embeddings. The cosine similarity between embeddings coming from two videos are then used for verification at test time. In this section, we will describe the uni-modal and multi-modal systems that allow us to generate these embeddings. 

Our \textit{unimodal} systems consist of an encoder ($F$) followed by a nonlinear classifier ($\mathcal{C}$) as shown in Fig. \ref{fig:unimodala} and \ref{fig:unimodalv}. Therefore, the final network output is represented by 
\begin{align}
    y_i = \mathcal{C}_i (F_i(\mathbf{x}_i)) \quad i \in \{A, V\}
\end{align}
where the subscript $i$ in $\mathbf{x}_i$ denotes the modality of the input. In order to achieve AV speaker verification using unimodal systems, we use late score fusion. In this fusion, we separately compute the cosine similarity using each unimodal system and then average the similarities to get the final verification scores.



For joint AV training, we investigate a naive \textit{mid-level fusion} approach. This is shown in Fig. \ref{fig:concat}. In this model, we have separate encoders for audio and video, whose outputs are concatenated along the feature dimension before being fed into a nonlinear classifier. The network output is represented as   
\begin{align}
    y_{AV} = \mathcal{C}_{AV} ([F_A(\mathbf{x}_A), F_V(\mathbf{x}_V)]).
\end{align}
In this case, a single loss function is applied to the joint output $y_{AV}$ during training. During verification, AV embeddings from two AV inputs are compared using cosine similarity. 

\section{The Multiview Model}
\label{sec:mview}
We propose a model that is trained to generate high level representations for audio and video modalities in a space shared across the two modalities. Such a system enables us to use the learned embeddings in a cross-modal testing scheme. We achieve this by using a shared classifier for audio and video encoder outputs and hence when optimized jointly, the encoder outputs are mapped to a shared space. We call this system a multi-view system since the classifier sees different views of the same input, i.e. the audio component and the visual component of the video input. 

As shown in Fig.~\ref{fig:multiview}, in the multi-view model, we still have two separate encoders for audio and video ($F_A $ and $F_V$). If we denote the multi-view classifier by $\mathcal{C}_{M}$, then the network will have two outputs one for each modality: 
\begin{align}
    y_{M,A} &= \mathcal{C}_{M} (F_A(\mathbf{x}_A))\\
    y_{M,V} &= \mathcal{C}_{M} (F_V(\mathbf{x}_V)).
\end{align}

In this study, we jointly train the whole network with a multi-task objective. The total loss $\mathcal{L}$ is calculated using 
\begin{align}
    \mathcal{L}_{M,AV} = \lambda_{M,A} \mathcal{L}_{M,A} + \lambda_{M,V} \mathcal{L}_{M,V}
\end{align}
where the unimodal losses $\mathcal{L}_A$ and $\mathcal{L}_V$ are computed based on $y_{M,A}$ and $y_{M,V}$, respectively. Note that here we jointly optimize two encoders and a single shared classifier. 

\section{Experiments and Results}
\label{sec:exps}
We perform our training on VoxCeleb2 (VC2) dev set and test on both VoxCeleb1 (VC1) and VC2 test sets. For the VC1 case, we use the verification pairs provided as part of the dataset. For VC2, we sample one positive and one negative video for each test set video as there is not an official set of pairs provided with the dataset.
The positive video is randomly uniformly chosen from the utterances of the same speaker and the negative video is sampled from a different speaker by first choosing a random speaker and then selecting a random utterance of that sampled negative speaker.  
To get the training and validation splits, we set aside one video (including several utterances) of each speaker for validation and use the rest for training. This gives us roughly 995k training utterances and 97k validation utterances with their corresponding visuals. 

\vspace{-9pt}
\subsection{Implementation Details}
We use 64-dimensional logmel features to represent the audio. For the video component, we downsample the data to 2 frames per second and apply face detection to each frame using Detectron2 \cite{wu2019detectron2}, then resize the face crops into 112x112 pixels. We skip the frames for which we do not have a face detection output. If we cannot detect any of the faces in a video, we use a single zero frame to represent it.  

In our \textit{unimodal} and \textit{mid-level AV fusion} systems, we make use of variants of convolutional neural networks (CNNs) for modeling the encoders. For the audio-only network and for the audio branch of the AV network, we use MobileNetV2 \cite{sandler2018mobilenetv2} with the following inverted residual blocks dimensions: [3,32,1,1], [4,32,1,1], [6,64,1,2], [4,64,1,1], [4,64,1,1], [6,128,1,2], [6,128,1,1], [4,128,1,1], [6,256,1,2], [4,256,1,1], [5,256,1,1], [4,256,1,1] based on the notation used in \cite{sandler2018mobilenetv2}. For the video-only branch of the AV model, we use ResNet architecture \cite{he2016deep}. At the end of encoders, we have a sequence of features on both branches. In order to summarize the utterance into a single vector we pool using a self-attention mechanism on the audio-branch and temporal pooling on the video branch. These result in 356 dimensional audio encodings and 2048 dimensional video encodings. In the multi-view system, in order to bring them into the same dimensions, in our case 256, we apply a linear projection layer on both branches before feeding them into the classifier.  
The classifiers consist of a fully connected layer, followed by ReLU nonlinearity, batch normalization \cite{ioffe2015batch}, dropout \cite{srivastava2014dropout} and a linear fully connected layer.

We train all the networks with arc-margin loss \cite{deng2019arcface}. 
We use a learning rate of 0.001 which is reduced by a factor of 0.95 based on the plateau of the loss value, the batchsize is 128. 




\vspace{-9pt}
\subsection{Experiments on Unimodal and Multimodal Models}
In Table~\ref{table:uni-naive-eer}, we present the EERs on VC1 and VC2 datasets.
The upper part of the table includes audio-only (A-only) EER of \cite{yadav2020frequency,zeinali2019but} and attention based fusion proposed in \cite{shon2019noise}. In the lower part of the table, the first two rows show the \textit{unimodal} performance of our systems. In the case of A-only, we achieve comparable results to the current best performance on VC1. Although \cite{zeinali2019but} reports lower EER, they make use of either heavy data augmentation or system fusion. The third row reports EER for our \textit{mid-level fusion} approach. Since it makes use of both modalities, the EER is lower than either of the \textit{unimodal} systems. We also experimented with score fusion by averaging the cosine similarity scores from various systems before making the verification decision. The late fusion of \textit{unimodal} systems achieves even lower performance than the naive AV fusion possibly because separately optimized A-only and V-only systems learn to capture the best representation of their respective input and their late fusion allows combining the best decisions from each modality. Furthermore, if we combine all three systems, then we achieve the lowest EERs on both VC1 and VC2. Note that our VC2 EER is not directly comparable to the one reported in \cite{shon2019noise} as we are not using the same verification pairs. Still, we achieve the lowest EER that has been reported on the VC2 test set so far. 

On manually cross checking the results of our system with the labels in the VC2 test set, we observe some interesting error cases. First, there is a small percentage (estimate 1\%) of the dataset which is labelled incorrectly. Second, there are cases where a person consciously obfuscates their voice or face (by wearing makeup or deliberately changing their voice), leading to hard cases where there is a deliberate mismatch between this atypical sample and videos where the person is acting naturally.
\begin{table}[t]
\centering
\begin{tabular}{ |l|c|c| } 
 \hline
Model description & VC2 EER & VC1 EER\\
 \hline
 A-only of \cite{zeinali2019but} & NA & 1.0 \\
 A-only of \cite{yadav2020frequency} & NA & 2.0  \\ 
 AV of \cite{shon2019noise} & 5.3 & NA 
 \\ \hline \hline
Our unimodal A-only & 3.5 & 2.2 \\
Our unimodal V-only & 3.4 & 3.9 \\
Mid-level AV fusion & 2.0 & 1.4 \\
 Score fusion unimodal A+V    & 1.7 & 0.9 \\
 Score fusion A+V+AV  & \textbf{1.6} & \textbf{0.7}\\
 \hline
\end{tabular}
\caption{EER (\%) of various models on VC1 and VC2 test sets}
\label{table:uni-naive-eer}
\vspace{-15pt}
\end{table}

\vspace{-9pt}
\subsection{Experiments on the Multi-view Model}
The \textit{multi-view} model allows for greater flexibility during testing as compared to the other models. With a \textit{multi-view} model, in addition to unimodal testing, we can apply late fusion to audio and video similarity scores, as well average audio and video embeddings to compute the similarity scores.

In Table \ref{tab:score-fusion-av}, we  show the \textit{unimodal} performance of the \textit{multi-view} model as well as their score fusion. 
When we compare the A-only performance of the \textit{multi-view} model with that of \textit{unimodal} model, we see some degradation in performance. However, the V-only performance of the \textit{multi-view} model is comparable to the \textit{unimodal} system, the differences are less than 0.2\%. We think that there are a couple of reasons for the reduction in the A-only performance: a) The \textit{multi-view} model requires that the intermediate dimensions of Audio and Video embeddings to be the same which are different than our unimodal systems and this causes reduction in the total number of trainable parameters, b) especially, at the beginning of training, statistics of audio and video embeddings differ so we had to remove the shared batch normalization layer from the classifier part of the network. That might have as well affected the performance. 
On the other hand, when we look at the score fusion of the \textit{multi-view} system given in last row of Table~\ref{tab:score-fusion-av}, we see that its EER is lower than either of the \textit{unimodal} systems reported in Table \ref{table:uni-naive-eer} (A-only, V-only). This also shows that score fusion is a simple but an effective mechanism to reduce the EER. Another observation that we make is that it is harder to optimize both embeddings simultaneously as compared to the separate training case which causes the EER difference between A-only and V-only test cases as compared to Table \ref{table:uni-naive-eer}.


As described in Section \ref{sec:mview}, the main goal of the \textit{multi-view} model is to do cross-modal testing. We simulate this A vs. V testing condition by dropping the audio modality from one side and the video modality from the other side of the verification pair. Another critical point is that both VC2 and VC1 test speakers are unheard and unseen during training. This makes the problem challenging as we try to match the face of a previously unseen person to the voice of a previously unheard person. It has been shown that even human performance on this task is low (more than 20\% error \cite{smith2016matching}). Since A vs. V cross-modal verification setting is the most difficult situation, it has higher EER as compared to Table~\ref{tab:score-fusion-av} but it is still better than the 50\% chance level. Table~\ref{table:cross-modal} shows cross-modal EER of our system and other published systems on VC1 and VC2 test sets. Here we observe that our system's performance is comparable to that of previously published systems. However, we cannot claim that our system is better or worse as the other works do not use the VoxCeleb dev/test splits in a similar manner.

We also performed A vs. AV and V vs. AV type of verification tests which are lacking one modality only on one side. Our experiments show that in such scenarios, it is better to use the matched data (A vs. A, or V vs. V) rather than fusing audio and video embeddings linearly, i.e. taking the average of audio and video embeddings. This is probably because of the fact that the shared space is not a linear space and it does not necessarily cover the linear combination of two embeddings. 


\begin{table}[t]
    \centering
    \begin{tabular}{|l|c|c|}
    \hline
      Test condition & VC2 EER & VC1 EER \\ 
      \hline
     Multi-view A-only & 7.2 & 6.1 \\
     Multi-view V-only & 3.5 & 3.7 \\
     Multi-view Score fusion A+V & 2.4 & 1.8 \\
      \hline
    \end{tabular}
    \caption{Audio-only, video-only and score fusion results from the multi-view system on both VC1 and VC2}
    \label{tab:score-fusion-av}
\end{table}

\begin{table}[t]
\centering
\begin{tabular}{ |l|c|c| } 
 \hline
Test pairs & VC2 EER & VC1 EER \\
 \hline
 A vs V of \cite{nagrani2018learnable} & NA & 29.5 \\
 A vs V of \cite{nawaz2019deep} & NA & 29.6 \\
 A vs V of \cite{tao2020audio} & 22.5 & NA \\ \hline \hline
 A vs V (our)  & 29.5 & 28.0 \\   
 \hline
\end{tabular}
\caption{Cross-modal verification EER on VC1 and VC2 test sets from previous studies and the proposed multi-view model}
\label{table:cross-modal}
\vspace{-12pt}
\end{table}

\section{Conclusions}
\label{sec:concl}
In this work, we first investigated AV speaker verification on VoxCeleb datasets. We learned the AV embeddings from VC2 dataset and then applied cosine similarity based verification on both VC2 and VC1 test sets. We showed that with score fusion of \textit{unimodal} and \textit{mid-level AV fusion} models, we achieve the lowest EER reported on VC1 test set in the AV testing condition. We also proposed a multi-view system that maps audio and video to a shared space and enables the cross-modal verification scenario of real verification systems. 


\newpage

\begin{spacing}{0.9}
\bibliographystyle{IEEEbib}
\bibliography{strings,refs}
\end{spacing}

\end{document}